\documentclass[11pt,tightenlines,prd,amsmath,amsfonts,amssymb,nofootinbib,floatfix]{revtex4}

\usepackage{epsfig}

\newcommand{\be}{\begin{equation}}
\newcommand{\ee}{\end{equation}}
\newcommand{\bal}{\begin{align}}
\newcommand{\eal}{\end{align}}
\newcommand{\ba}{\begin{eqnarray}}
\newcommand{\ea}{\end{eqnarray}}
\newcommand{\bi}{\begin{itemize}}
\newcommand{\ei}{\end{itemize}}
\newcommand{\bc}{\begin{center}}
\newcommand{\ec}{\end{center}}
\newcommand{\bfig}{\begin{figure}}
\newcommand{\efig}{\end{figure}}

\renewcommand{\vec}[1]{{\bf {#1}}}

\newcommand{\as}{\bar\alpha}

\newcommand{\xgg}{\chi^{\prime\prime}}
\newcommand{\xgw}{\dot\chi^{\prime}}
\newcommand{\xg}{\chi^{\prime}}
\newcommand{\xw}{\dot\chi}
\newcommand{\xww}{\ddot\chi}
\newcommand{\x}{\chi}

\newcommand{\g}{\gamma}
\newcommand{\w}{\omega}
\newcommand{\gc}{\gamma_c}
\newcommand{\wc}{\omega_c}
\newcommand{\dL}{\partial_L}
\newcommand{\dY}{\partial_Y}
\newcommand{\dt}{\partial_t}
\newcommand{\dx}{\partial_x}

\newcommand{\eq}[1]{(\ref{#1})}

\begin{document}

\title{Traveling waves and the renormalization group improved\\ Balitsky--Kovchegov equation}

\author{Rikard Enberg}
\email[Email:\ ]{REnberg@lbl.gov} 
\affiliation{Theoretical Physics Group, Lawrence Berkeley National Laboratory, Berkeley, CA 94720, USA}

\begin{abstract}
I study the incorporation of renormalization group (RG) improved BFKL kernels in the Balitsky--Kovchegov (BK) equation which describes parton saturation. The RG improvement takes into account important parts of the next-to-leading and higher order logarithmic corrections to the kernel. The traveling wave front method for analyzing the BK equation is generalized to deal with RG-resummed kernels, restricting to the interesting case of fixed QCD coupling. The results show that the higher order corrections suppress the rapid increase of the saturation scale with increasing rapidity. I also perform a ``diffusive'' differential equation approximation, which illustrates that some important qualitative properties of the kernel change when including RG corrections.
\end{abstract}

\vspace{-20cm}

\hfill LBNL-61909

~

\maketitle

\section{Introduction}

Studies of parton saturation and the unitarization of high energy amplitudes began more than 20 years ago with the work of Gribov, Levin and Ryskin \cite{GLR} and has now evolved into a very active field of research \cite{reviews}. The simplest modern QCD evolution equation that describes parton saturation in the high energy limit is the Balitsky--Kovchegov (BK) equation~\cite{Kovchegov1,Balitsky}. The BK equation is a ``mean-field'' approximation which neglects higher order correlators present in the so-called Balitsky \cite{Balitsky} and JIMWLK equations (see \cite{reviews} and references therein). Nevertheless, because the corrections neglected in the mean field approximation are likely small at phenomenologically interesting energies \cite{EGM}, it is expected to be accurate for dipole scattering on large nuclei and at least a phenomenological model for saturation in scattering on protons. The equation has the correct high-energy limit at small parton densities, when it reduces to the BFKL equation \cite{BFKL}, but it has a non-linear term which reduces the power-like growth with energy.

The BK equation is derived at leading logarithmic (LL) accuracy and thus reduces to the LL BFKL equation in the linear regime. The next-to-leading logarithmic (NLL) corrections to the BFKL equation were computed in \cite{NLLBFKL} and were found to be negative and very large. As the BFKL kernel is a central part of the BK equation, it is important to understand the effect of the NLL corrections on the solutions of the BK equation.

In fact the NLL BFKL kernel is larger than the leading order kernel, and for any reasonable value of $\alpha_s$ it is negative; the pomeron intercept becomes less than one for $\alpha_s\gtrsim 0.15$. The kernel also has two complex conjugate saddle points which lead to oscillating cross sections. The perturbative expansion of the BFKL kernel is therefore highly unstable and far from converging. These problems are discussed in detail in \cite{NLLproblems}.

A cure was proposed by Ciafaloni, Colferai, and Salam \cite{CCS}, who realized that the large corrections come from collinear contributions of the DGLAP type. They devised the so-called $\omega$-resummation, which makes the NLL kernel compatible with renormalization group (RG) requirements through matching to the DGLAP limit and resummation of spurious poles. A somewhat different scheme has been proposed in \cite{ABF}. Many groups have also studied ways of including some of the most important NLL effects in the LL kernel \cite{cconstraint,rapveto,Khoze:2004hx}. A common feature of these schemes is that they modify the functional dependence of the Mellin space BFKL kernel $\chi(\gamma)$ so that it acquires a dependence also on the $\omega$ variable. This will be discussed at length below. In a slight abuse of terminology I will henceforth refer to all these schemes as ``RG improved'' kernels.

Another important part of the NLL corrections is the running of the QCD coupling. It is, however, not clear how to incorporate the effect of the running coupling in the BK and Balitsky--JIMWLK equations, and recent studies have come to different conclusions \cite{running}. It is also reasonable, but by no means obvious, to assume that the NLL corrections to the BFKL part of the BK equation will be the same as for the BFKL equation. Since the way to include the running coupling is unclear it is therefore useful to study the BK equation with NLL corrections to the BFKL kernel but with a fixed coupling.

The onset of saturation corrections is governed by the saturation scale $Q_s$---the momentum that separates the regions dominated by linear evolution and saturation. Because of the non-linear term in the BK equation it is difficult to obtain analytical solutions, but at large momenta the amplitude is small enough that one may consider the linear equation only (the BFKL equation) together with appropriate boundary conditions. This has been used in a series of papers \cite{Mueller:1999wm,IancuItakuraMcLerran,Dionysis1,Dionysis2} to compute the rapidity dependence of $Q_s=Q_s(Y)$. Recently the method of traveling waves has been developed~\cite{MP1,MP2,MP3}, which allows performing these calculations in a more systematic way.
Asymptotically the saturation scale increases exponentially as rapidity increases \cite{Mueller:1999wm,IancuItakuraMcLerran}, but, as we will discuss below, there is also a factor with a sub-asymptotic $Y$-dependence which slows down the increase \cite{Dionysis1,Dionysis2,MP1}.

Saturation is intimately related to geometric scaling, which was discovered in the HERA data in Ref.\ \cite{geometricscaling}. It refers to a scaling of the $\gamma^* p$ cross section for small Bjorken-$x$, where the cross section $\sigma^{\gamma^* p}(x,Q^2)$ is not a function of $x$ and $Q^2$ separately, but only through the combination $\tau=Q^2/Q_s^2(x)$.
As we shall see, the BK equation predicts geometric scaling of the amplitude and a specific pattern of violation of this scaling.

In this paper I make an analytical study using the method of traveling wave fronts to compute the higher-order corrections and suggest a method which is simple to implement for any higher order corrected kernel (see also \cite{RE-DIS2005}). I will do this for the case of a fixed strong coupling $\as$. The case of an RG-improved kernel with running coupling was very recently done by Peschanski and Sapeta \cite{PS} and will therefore not be studied here. The present paper should therefore be taken together with \cite{PS} for a treatment of the RG-improved BK equation.

The outline of the paper is as follows. Section \ref{II} contains a brief review of the traveling wave front method for analyzing the BK equation and section \ref{III} discusses higher order corrections to the BFKL kernel. In Section \ref{IV},  I then generalize the method described in Section \ref{II} to deal with RG-resummed kernels. I show how to compute analytic results for the saturation scale and the shape of the solution in momentum space above the saturation scale, restricting to the case of fixed coupling. This is developed for a general form of the resummed kernel and then applied to various schemes proposed in the literature in Section \ref{V}. In Section \ref{VI}, finally, I perform a ``diffusive'' differential equation approximation, which illustrates that some important qualitative properties of the kernel change when including RG corrections.

\section{The BK equation and traveling waves} \label{II}

The BK equation in position space 
describes the scattering amplitude ${\cal N}(\vec x_{01},Y)$ of a dipole on a target (originally a large nucleus), where the target is specified by the initial condition. For large and homogeneous targets (target radius much larger than the involved dipoles in the cascade) one may neglect the dependence on the impact parameter and Fourier transform to momentum space. The equation can then be written on the form~\cite{Kovchegov2}
\be
\dY {\cal N} = \as \chi_0(-\dL) {\cal N} - \as {\cal N}^2.
\label{LLBK-eq-fixed}
\ee
where $\as=N_c \alpha_s/\pi$, $L=\ln(k^2/k_0^2)$, with $k_0$ some low momentum scale, e.g., $\Lambda_{QCD}$, and $\dY\equiv\partial/\partial Y$, $\dL\equiv\partial/\partial L$. The quantity ${\cal N}(\vec k,Y)$ which we will study in the following is related to the unintegrated gluon distribution of the target. The non-linear term in Eq.\ \eq{LLBK-eq-fixed} is particularly simple in this representation and is not acted on by the kernel operator.
Linearizing Eq.\ \eq{LLBK-eq-fixed} one arrives at the BFKL equation \cite{BFKL}.

The operator $\chi_0(-\dL)$ is given by the characteristic BFKL function 
\be
\chi_0(\gamma) = 2\psi(1) - \psi(\gamma) - \psi(1-\gamma),
\label{chi0}
\ee
the Mellin transform of the LL BFKL kernel in momentum space \cite{BFKL}. The explicit form of the term $\chi_0(-\dL) {\cal N}$ is a convolution $K_\text{BFKL}\otimes{\cal N}$ of the BFKL integral kernel $K_\text{BFKL}$ and the gluon distribution, but it can be formally represented as an infinite order differential operator through the power series
\be
\chi_0(-\dL) {\cal N} = \left[ \chi_0(\g_0) 
+ (-\dL-\g_0) \chi_0'(\g_0)
+ \frac{1}{2!}(-\dL-\g_0)^2 \chi_0''(\g_0) +\dots \right] {\cal N}
\label{series-kernel}
\ee
where $\g_0$ is a constant. 

The approximations involved in deriving the BK equation are (i) large $N_c$, (ii) no correlations between different dipoles,  and (iii) leading logarithmic (LL) resummation. In this paper we will investigate the corrections to the third point. The second point has been studied a lot in the last years, e.g., in \cite{Iancu:2004es,EGM}.

Munier and Peschanski~\cite{MP1,MP2,MP3} found an interesting relationship between non-linear evolution equations in QCD (in particular the BK equation) and statistical physics studies of \emph{traveling wave fronts}. These are wave fronts that separate a stable state from an unstable state, which invades the unstable state by propagating in space with an essentially fixed shape. Such fronts are described by non-linear diffusion equations with a growth term and a reduction term \cite{Fisher,KPP,travelingwaves,EvS}. 

They observed~\cite{MP1} that if one truncates the series \eq{series-kernel} and keeps only the terms up to order $\dL^2$ one obtains the differential equation
\be
\dY {\cal N} = \as \left[ \chi_0(\g_0) 
+ (-\dL-\g_0) \chi_0'(\g_0)
+ \frac{1}{2!}(-\dL-\g_0)^2 \chi_0''(\g_0)\right] {\cal N} - \as {\cal N}^2
\label{LLBK-KPP}
\ee
which, with a suitable change of variables $Y \propto t$ and $L = A x + B t$ (with $A,B$ some constants), gives the equation
\be
\dt u = \dx^2 u +  u -  u^2.
\label{FKPP}
\ee
This is the celebrated Fisher--Kolmogorov--Petrovsky-Piscunov
(FKPP) equation~\cite{Fisher,KPP}, which under certain conditions leads to  traveling wave solutions $u=u(x,t)$.
The condition is that the initial conditions are steep enough, i.e.\ decay like $\exp(-\lambda x)$ at large $x$ with $\lambda$ large enough, see~\cite{travelingwaves,EvS} for more details. I will not discuss the initial conditions further here, but will assume that they are steep enough, as is the case in QCD~\cite{MP1}.

The traveling wave property means essentially that the solutions are saturated wave fronts in the space variable $x$, which propagate forward in space as the time $t$ increases, with a fixed (or nearly fixed) shape. The speed of propagation $v(t)$ of this front has been known for a long time~\cite{Bramson}; it approaches an asymptotic speed $v^*$ like
\be
v(t) = v^* - \frac{3}{2 \lambda^* t} + {\cal O}(t^{-3/2})
\label{velocity2}
\ee
where for the FKPP equation $v^*=2$ and $\lambda^*=1$. The result for the saturation scale given in~\cite{MP1} is obtained from this formula by using the correct change of variables to get the approximate BK equation \eq{LLBK-KPP} in the form of the FKPP equation, Eq.\ \eq{FKPP}. That is, when studying the BK equation in the traveling wave framework, we study the evolution of a front that propagates in the ``space'' variable $L\sim\ln k^2$ as the ``time'' $Y$ increases.

The $O(t^{-3/2})$ term in the velocity relation for FKPP-like equations has been found by Ebert and van Saarloos~\cite{EvS}, namely
\be
v(t) = v^* - \frac{3}{2 \lambda^* t}\left( 1 - \sqrt{\frac{\pi}{(\lambda^*)^2 D t}}\right)  + {\cal O}(t^{-2})
\label{velocity}
\ee
where $\lambda^*$ is related to the wave number and has the dimensions of inverse length, and $D$ is the diffusion constant. (For the FKPP equation \eq{FKPP}, $D=1$.) It has also been shown that these are all universal terms in the velocity relation, i.e., all other terms depend on the initial conditions. The three constants $v^*$, $\lambda^*$ and $D$ \emph{completely determine} the velocity relation of the traveling waves, and thus the saturation scale in the case of the BK equation. These constants are calculated from the dispersion relation for Fourier modes of the solutions of the equation under study.

The most important point of the approach of Munier and Peschanski may be that the velocity that one calculates is \emph{independent} of the exact form of the non-linearity, i.e., it is enough to analyze the linearized integro-differential equation, i.e., the BFKL equation, which of course is much easier than the full BK equation. This does not mean that one gets saturation from the linear equation---the non-linearity is crucial in selecting the velocity \eq{velocity}. See Section III of Ref.\ \cite{EvS} for a thorough discussion of this point.

\section{Higher order corrections to the BFKL and BK equations} \label{III}

The full NLL BFKL kernel is given by $\chi(\gamma)=\chi_0(\gamma)+\as \chi_1(\gamma)$ where $\chi_0$ is the LL BFKL kernel and $\chi_1$ is the second order contribution to the BFKL kernel and is given in Refs.\ \cite{NLLBFKL}. The full BK equation has not been derived at NLL  accuracy---only the BFKL equation is known. There will also be corrections to the non-linear term, but it is reasonable to assume that the linearized equation will be the same as the NLL BFKL equation. Since the observables studied here are, as explained above, determined by the linear equation we can study the BK equation with NLL corrections only in the linear term and obtain the correct $Y$-dependence of the saturation scale. The absolute normalization of $Q_s$ is not constrained by the method. The normalization of the wave front, however, is given by the full non-linear evolution of the initial condition and may change if the non-linear term is changed. The same holds for the wave front in the saturation region, but here we only aspire to calculating it in the region above the saturation scale.

Using the improved kernels mentioned in the Introduction, the BFKL kernel is modified to the form $\chi(\gamma,\omega)$ depending on both $\gamma$ and $\omega$. 
The RG-improved BK equation is then non-local in both $L$ and $Y$ and mixes transverse and longitudinal degrees of freedom (see also Section \ref{diffsection}). It can be written
\be
\dY {\cal N} = \as \chi(-\dL,\dY) {\cal N} - \as {\cal N}^2
\ee
regardless of the specific form of the kernel. Here one can take either fixed or running coupling. The evolution equation is now non-local in rapidity, whereas the uncorrected kernel (LL or NLL) only is non-local in momentum. This is where the traveling wave method reveals its power---it can easily be generalized to kernels much more complicated than LL BFKL (and similarly the FKPP equation can be turned into higher order differential equations). 

The first study of higher order corrections to the saturation scale was to my knowledge done by Golec-Biernat et al.\ \cite{Golec-Biernat:2001if} in their numerical study of the BK equation, by implementing the kinematical constraint of Ref.\ \cite{cconstraint}. The first analytical study was performed by Triantafyllopoulos \cite{Dionysis2} using BFKL evolution with absorbing boundary conditions \cite{Dionysis1}. The higher corrections in this case were taken into account through the use of the resummed NLL kernel of Refs. \cite{CCS} with a running coupling constant. Khoze et al.\ \cite{Khoze:2004hx} have made another BFKL study where they compute the asymptotic term in the saturation scale using a resummed LL kernel along the lines of~\cite{CCS}, i.e., a simpler variant than the one in \cite{Dionysis2}. Chachamis et al.\ \cite{Chachamis:2004ab} used the BFKL kernel with a rapidity veto \cite{rapveto} in the BK equation and performed a numerical solution of this BK equation. Finally, Gotsman et al.\ \cite{Gotsman:2004xb} studied a DGLAP-corrected BFKL kernel similar to the one used in \cite{Khoze:2004hx}. All these studies naturally find a reduced growth with energy of the saturation momentum.

\section{Method for calculation} \label{IV}

\subsection{The leading logarithmic BK equation}\label{LL-BFKL-fixed}

The linearized equation corresponding to \eq{LLBK-eq-fixed} is the LL BFKL equation, which can be solved analytically by Mellin transformation; the solution is given by the double inverse Mellin transform\footnote{Note that I use a slightly different convention than e.g.~\protect\cite{MP1} for the $\gamma$-dependence of the exponent.}
\be
{\cal N}(k,Y) = \int_{\cal C} \frac{d\gamma}{2\pi i} \int_{{\cal C}^\prime} \frac{d\omega}{2\pi i} {\cal N}_0(\gamma,\omega) 
\frac{e^{-(1-\gamma) L +\omega Y}}{\omega - \as \chi_0(\gamma)}
\ee
where ${\cal N}_0(\gamma,\omega)$ encodes the initial conditions. This is analogous to the Fourier--Laplace transform used in \cite{EvS} to analyze the FKPP equation leading to a dispersion relation between wave number and frequency for the Fourier modes\footnote{Working with Mellin transforms instead of Fourier--Laplace transforms, the quantities $\omega$ and $k^*$ are real instead of purely imaginary. Appropriate changes must therefore be made to the formulas of \protect\cite{EvS}.}. In the present case we obtain a dispersion relation for Mellin moments by picking up the pole in the denominator to perform the $\omega$-integral. This gives
\be
{\cal N}(k,Y) = \int_{\cal C} \frac{d\gamma}{2\pi i} \widetilde {\cal N}_0(\gamma)
e^{-(1-\gamma) L + \as \chi_0(\gamma) Y}.
\ee
We thus have the dispersion relation $\omega(\gamma) = \as \chi_0(\gamma)$ for the solution of the BK equation. The evolution variable is now rapidity $Y$, which corresponds to time $t$ in the FKPP equation.

Let us now see how easily the saturation scale is computed using the Ebert--van Saarloos (EvS) method \cite{EvS}. The calculation consists of finding the three parameters $v^*$, $\lambda^*=k^*=1-\g_c$, and $D$.
The linearized LL-BK equation (i.e., the LL-BFKL equation) is in Mellin $\gamma$-space written as
\be
\dY \widetilde\phi(\gamma,Y) = \as \chi_0(\gamma) \widetilde\phi(\gamma,Y)
\ee
where $\widetilde\phi(\gamma,Y)$ denotes the Mellin transform of ${\cal N}(k,Y)$ with respect to $L$. 
Following the general formalism of EvS Section V-D, we write this as
\be
\left[\dY + \widetilde{\mathsf T}(\gamma)\right] \widetilde\phi(\gamma,Y) = 0
\label{eqn}
\ee
where $-\widetilde{\mathsf T} \widetilde\phi$ is the Mellin transform with respect to $k^2$ of $K_\text{BFKL}\otimes{\cal N}$, i.e., 
$\widetilde{\mathsf T} = -\as \x_0(\g)$.
Define also $\hat{\mathsf S}(\gamma,\omega)$, the Mellin transform with respect to $Y$ of the left hand side of \eq{eqn}, i.e., it is the double Mellin transform of the BFKL equation.
In general, for differential equations, $\hat{\mathsf S}$ and $\widetilde{\mathsf T}$ are $M \times M$ matrices where $M$ is the order of the highest time derivative.

The dispersion relation for Mellin moments is obtained from the requirement $S(\g,\w)=0$ where $S$ is the characteristic function $S(\gamma,\omega) = \det \hat{\mathsf S}(\gamma,\omega)= \omega -\as \x_0(\g)$. This again yields
$\omega(\gamma) = \as \chi_0(\gamma)$
as expected. One now finds the asymptotic velocity $v^*$ and the saddle point $k^* = \lambda^* = 1-\gc$ by solving the set of equations [EvS Eqs.\ (5.60) and (5.66)]
\be
\begin{cases}
\displaystyle v^* = -\left. \frac{d\omega(\gamma)}{d\gamma}\right|_{\gc} = -\as \xg_0(\gc)\\
\displaystyle v^* = \frac{\omega(\gc)}{1-\gc} = \frac{\as \x_0(\gc)}{1-\gc}.
\end{cases}
\label{gammac-eqs}
\ee
The relevant saddle point for evolution with constant amplitude is therefore not the usual BFKL saddle point $\gamma_0=1/2$ but is given by the solution of the equation $\xg_0(\gc)(1-\gc)+\x_0(\gc)=0$ as was shown already in \cite{GLR}. A numerical solution of \eq{gammac-eqs} gives $\gc=0.373$. 
Finally, the diffusion coefficient $D$ is obtained from
\be
D = \left.\frac{1}{2!} \frac{d^2\omega(\gamma)}{d\gamma^2}\right|_{\gc}
= \frac{\as}{2} \xgg_0(\gc).
\label{D}
\ee
The speed of the front is now given by Eq.\ \eq{velocity}
and the position is then simply
\be
x(Y)= \ln [Q_s^2(Y)/k_0^2]	 =  v^* Y - \frac{3}{2\lambda^*} \ln Y
- \frac{3}{(\lambda^*)^2} \sqrt{\frac{\pi}{D Y}} 
\ee
which, inserting $v^*$, $\lambda^*$and $D$ gives the result (neglecting $k_0$)
\be
\ln Q^2_s(Y) = \as \frac{\x_0(\gc)}{1-\gc} Y - \frac{3}{2(1-\gc)} \ln Y - \frac{3}{(1-\gc)^2} \sqrt{\frac{2\pi}{\as\xgg_0(\gc)}} \frac{1}{\sqrt{Y}}
+{\cal O}(1/Y).
\label{Qs-LL-fixed}
\ee
This expression for the saturation scale of a system whose evolution is described by Eq.\ \eq{LLBK-eq-fixed}
was first obtained in \cite{MP3}. These three terms are universal. The first term, which gives the asymptotic exponential rise of the saturation scale used e.g.\ in the saturation model of~\cite{GBW}, was obtained already in \cite{GLR} from the double logarithmic approximation, and from LL BFKL in \cite{IancuItakuraMcLerran}.
The second term was obtained in \cite{Dionysis1} using BFKL evolution with an absorbing boundary to mimic the non-linearities and was confirmed in \cite{MP1,MP2} using the traveling wave method.\footnote{Note that when comparing the results of \protect\cite{Dionysis1} and \protect\cite{MP1,MP2} there is a difference of a factor of 2 due to the convention for the definition of the BFKL characteristic function $\chi_0$. Furthermore, there are some terms in the results of \protect\cite{Dionysis1} that are non-universal and subleading in the classification used here.}

The shape of the wave front in the so-called leading edge region can also be computed in the  traveling wave framework. Ebert and van~Saarloos~\cite{EvS} compute the solution at the same level of approximation as the velocity \eq{velocity}, i.e., including all universal terms. Brunet and Derrida~\cite{BrunetDerrida} use a somewhat simpler method which allows computing the solution at the level of approximation of \eq{velocity2}, which gives simpler results. This method was employed in Ref.\ \cite{MP2} to compute the solution of the LL BK equation in the leading edge region for both fixed and running coupling. The result is
\begin{align}
{\cal N}(k,Y) &= \frac{1}{\sqrt{D}}
\ln\left(\frac{k^2}{Q_s^2(Y)}\right) 
\left(\frac{k^2}{Q_s^2(Y)}\right)^{\gc-1}
\exp\left(-\frac{1}{4 D Y}\ln^2\left(\frac{k^2}{Q_s^2(Y)}\right) 
\right)
\\
&= \sqrt{\frac{2}{\as\xgg_0(\gc)}}
\ln\left(\frac{k^2}{Q_s^2(Y)}\right) 
\left(\frac{k^2}{Q_s^2(Y)}\right)^{\gc-1}
\exp\left(-\frac{1}{2\as\xgg(\gc)Y}\ln^2\left(\frac{k^2}{Q_s^2(Y)}\right) 
\right) \nonumber
\end{align}
up to a constant factor. This result has been obtained both using an absorbing boundary~\cite{Dionysis1} and traveling waves~\cite{MP2}.

The agreement between these completely different methods both for the saturation scale and the wave front shape is good evidence that the traveling wave method is reliable and general. Furthermore, the numerical study of the BK equation in \cite{EGM} shows very good agreement at large $Y$ with Eq.\ \eq{Qs-LL-fixed} and the corresponding analytical formulas for ${\cal N}(k,Y)$ in the leading edge region.

The description in terms of operators $\hat{\mathsf S}$ and $\widetilde{\mathsf T}$ 
might seem unnecessarily formal, but a main result of the analysis of EvS~\cite{EvS} is that the method generalizes to much more general types of equations than differential equations of the FKPP type,\footnote{They have to be equivalent in the sense that they must be reaction--diffusion equation with second-order diffusion. The running coupling case, for example, leads to a different equivalence class which has $D=0$.} for example higher order differential equations, integral equations with a kernel in either time (memory kernels) or in space (non-local kernels), or even difference equations. 

Thus, the kernel in momentum or configuration space might have any complicated structure, but if one knows the Mellin (or Fourier) transform of the kernel, one can solve for the dispersion relation and compute $v^*, \lambda^*$ and $D$ (in fact, in the following treatment the higher order corrections are implemented in Mellin space rather than in momentum space). 
This dispersion relation can lead to non-algebraic equations, however, unlike for the FKPP equation. 
This generalization will be exploited below for the RG-resummed case.

\subsection{RG-improved BK}\label{NLL-BK-fixed}

Let me now generalize to RG-improved kernels at fixed coupling. The dispersion relation now reads
\be
\omega(\gamma,\as) = \as \chi(\gamma,\omega(\gamma,\as))
\label{NLLdisp}
\ee
which implicitly defines the function $\omega$. For each value of $\gamma$ 
and $\as$ Eq.\ \eq{NLLdisp} gives a transcendental equation, which can in principle be solved numerically for $\omega(\gamma)$. This is not what we will do here.

The asymptotic velocity $v^*$ and saddle point $\lambda^*=1-\gc$ are as earlier found from the equations
\begin{align}
v^* = & -\left. \frac{d\omega(\gamma,\as)}{d\gamma}\right|_{\gc}
\label{v1}\\
v^* = & \ \frac{\omega(\gc,\as)}{1-\gc},\label{v2}
\end{align}
but since $\omega(\gamma,\as)$ does not have an explicit form it is not as straightforward to obtain the required quantities. For LL there is a saddle point of the characteristic function $\x_0(\gamma)$ determined from the equation given above, but now the function $\x=\x(\gamma,\omega)$ has an additional variable. This means that we will have to determine two constants $\gc$ and $\wc$.
To keep notation simple, let us introduce the shorthand 
$\x_c = \chi(\gc,\wc)$, 
$\xg_c = \partial_\gamma \chi(\gc,\wc)$  and 
$\xw_c = \partial_\omega \chi(\gc,\wc)$.

First, we require $\wc = \as \x(\gc,\wc)$. Observing that on one hand, $v^*$ is given by
\be
v^* = -\left. \frac{d\omega}{d\gamma}\right|_{\gc} = -\as  \left.\left[
\frac{\partial\x(\gamma,\omega)}{\partial\gamma} + 
\frac{\partial\x(\gamma,\omega)}{\partial\omega}
\frac{d\omega}{d\gamma} \right] \right|_{\gc}, \label{dwdg}
\ee
so that 
\be
v^* = -\left. \frac{d\omega(\gamma)}{d\gamma}\right|_{\gc}
= -\frac{\as \xg_c}{1 - \as \xw_c},
\ee
and using Eq.\ \eq{v2} we obtain
$-\xg_c (1-\gc) = \x_c \, [1 - \as \xw_c] = \x_c  - \wc \,\xw_c$. We therefore have the system of equations
\be
\begin{cases}
\xg_c (1-\gc) + \x_c  - \wc \, \xw_c=0\\
\wc = \as \x_c
\end{cases}
\label{gcwc-eq}
\ee
The first equation reduces to the corresponding LL one if there is no dependence on $\omega$ in the kernel.

The advantage of knowing both $\gc$ and $\wc$ is obvious---there exists a closed form for all the constants that will appear in the expression for the saturation scale, and it is not necessary to use functions that can only be defined numerically and to numerically differentiate these functions. It is much more straightforward to just numerically solve \eq{gcwc-eq}.

Once we have determined $\gc$ and $\wc$, the other constants $v^*$ and $D$ are fixed (remember $\lambda^*=1-\gc$). We are then in a position to write down the expression for the velocity of the front,
\be
v(Y) = v^* - \frac{3}{2 (1-\gc) Y}\left( 1 - \sqrt{\frac{\pi}{(1-\gc)^2 D Y}}\right) 
\label{velocityY}
\ee
where $v^*=\wc/(1-\gc)$ and $D$ is obtained by applying the chain rule as in Eq.\ \eq{dwdg} twice;
\be
D = \frac{1}{2} \, \left. \frac{d^2\omega}{d\gamma^2}\right|_{\gc}
= \frac{\wc}{2\xg_c (\gc-1)}  
\left( \xgg_c + 2\, \frac{\wc}{1-\gc}\, \xgw_c 
+ \left(\frac{\wc}{1-\gc}\right)^2 \xww_c \right).
\label{diffusionD}
\ee
Finally this gives the saturation scale,
\be
\ln Q^2_s(Y) = \frac{\wc}{1-\gc} Y - \frac{3}{2(1-\gc)} \ln Y -  
\sqrt{\frac{18\pi \xg_c}{\wc (\gc-1)^3 Y}}
\left( \xgg_c + 2\, \frac{\wc}{1-\gc}\, \xgw_c 
+ \left(\frac{\wc}{1-\gc}\right)^2 \xww_c \right)^{-1/2}.
\label{Qs-NLL-fixed}
\ee

All the discussion so far is perfectly general in that we have not chosen any particular form for the NLL BFKL kernel, we have only anticipated that it will depend on both $\gamma$ and $\omega$.

\section{Application to RG-improved kernels}\label{applications}\label{V}

The method from the last section can now be easily applied to some of the RG-improved kernels suggested in the literature. I will look at the rapidity veto model of \cite{rapveto}, the kinematical constraint model of \cite{cconstraint}, and the resummed model of \cite{Khoze:2004hx}. I will thus solve Eqs.\ \eq{gcwc-eq} for these models numerically using Mathematica, and use the obtained values for $\gc,\wc$ in the expression \eq{Qs-NLL-fixed} for the saturation scale and in the generalized expression \eq{redfront} below for the shape of the wave front.

\subsection{Rapidity veto}
The idea of a rapidity veto in the kernel \cite{rapveto} is that by forbidding gluon emissions that are very close in rapidity one accounts for a part of the NLL corrections. This was studied numerically in the context of the BK equation in Ref.\ \cite{Chachamis:2004ab}.

The form of the kernel is 
\be
\x(\g,\w)=\x_0(\g) e^{-\eta \w}
\label{chi-rapveto}
\ee
where $\eta$ is the minimum rapidity interval imposed on two subsequent emissions, and is taken as a parameter. A numerical solution for $\omega(\gamma)$ is shown in Fig.\ \ref{CC-fig}.

The equations \eq{gcwc-eq} thus take the form
\begin{align}
\w_c = \ &  \as \x_0(\gc) e^{-\eta \w_c} \nonumber\\
\xg_0(\gc) \gc e^{-\eta \w_c} = \ &
\x_0(\gc) e^{-\eta \w_c} \left( 1 + \as \eta \x_0(\gc) e^{-\eta \w_c}\right)
\end{align}
which in fact for this factorized form of the kernel allows eliminating $\wc$ from the result so that one arrives at the following equation for $\gc$:
\be
\frac{\xg_0(\gc)(1-\gc)}{\x_0(\gc)}+1 = -\as \eta \x_0(\gc) \exp\left(1+\frac{\xg_0(\gc)(1-\gc)}{\x_0(\gc)}\right)
\label{gc-rapveto}
\ee
which, taking into account the difference in convention, $\gamma \to 1-\gamma$, is the result for $\gc$ found in Ref.\ \cite{Chachamis:2004ab}.
Note that for $\eta\to 0$ this reduces to the LL result. 

The authors of \cite{Chachamis:2004ab} give an analytic result for the saturation scale with only the asymptotic term for the velocity and then solve the BK equation numerically. But using the present method it is easy to find the subleading terms for the saturation scale. To do this we need to find the diffusion coefficient given by Eq.\ \eq{diffusionD}, 
\be
D =  \frac{\as \x_0(\gc)}{2\xg_0(\gc)(\gc-1)} e^{-\eta \w_c} 
\left( \xgg_0(\gc) - 2\eta\,\xg_0(\gc) \, \frac{\as \x_0(\gc)e^{-\eta \w_c}}{1-\gc}\, 
+ \eta^2 \, \x_c(\gc)\left(\frac{\as \x_0(\gc)e^{-\eta \w_c}}{1-\gc}\right)^2  \right)
\label{D-rapveto}
\ee
with
\be
e^{-\eta \w_c}=\exp \left(1+\frac{\xg_0(\gc)(1-\gc)}{\x_0(\gc) }\right).
\ee
Thus, the saturation scale with all universal terms reads
\be
\ln Q_s^2(Y) = \as \frac{\x_0(\gc)}{1-\gc} Y \exp \left(1+\frac{\xg_0(\gc)(1-\gc)}{\x_0(\gc) }\right)
- \frac{3}{2(1-\gc)} \ln Y
- \frac{3}{(1-\gc)^2} \sqrt{\frac{\pi}{D Y}} 
\ee
with $\gc$ and $D$ given by Eqs.\ \eq{gc-rapveto} and \eq{D-rapveto}.

\subsection{Kinematical constraint}

The so-called kinematical constraint, or consistency constraint \cite{cconstraint}, imposes a cut-off on the virtualities of gluons in the real emission part of the BFKL kernel, motivated by the requirement that in the high energy kinematics considered, the virtualities of the exchanged gluons should be dominated by their transverse momenta. It is also a part of the resummed framework of \cite{CCS}. 

This constraint has been shown to reproduce a large part of the NLL-corrections to the pomeron intercept. It has been used in several phenomenological applications of BFKL dynamics. Because the value of the pomeron intercept is reduced, the results show a reduced growth of cross sections with energy, and generally fit data better than LL BFKL.
 
The constraint implies shifting the $\gamma$ in the kernel by $\pm\omega/2$,
\be
\x(\g,\w)=2\psi(1)-\psi\left(\g+\frac{\w}{2}\right)-\psi\left(1-\g+\frac{\w}{2}\right),
\label{chi-CC}
\ee
such that, in contrast to the kernel \eq{chi-rapveto} with rapidity veto discussed above, the kernel \eq{chi-CC} does not have poles in $\g=0,1$.

\begin{figure}[th]
\epsfig{file=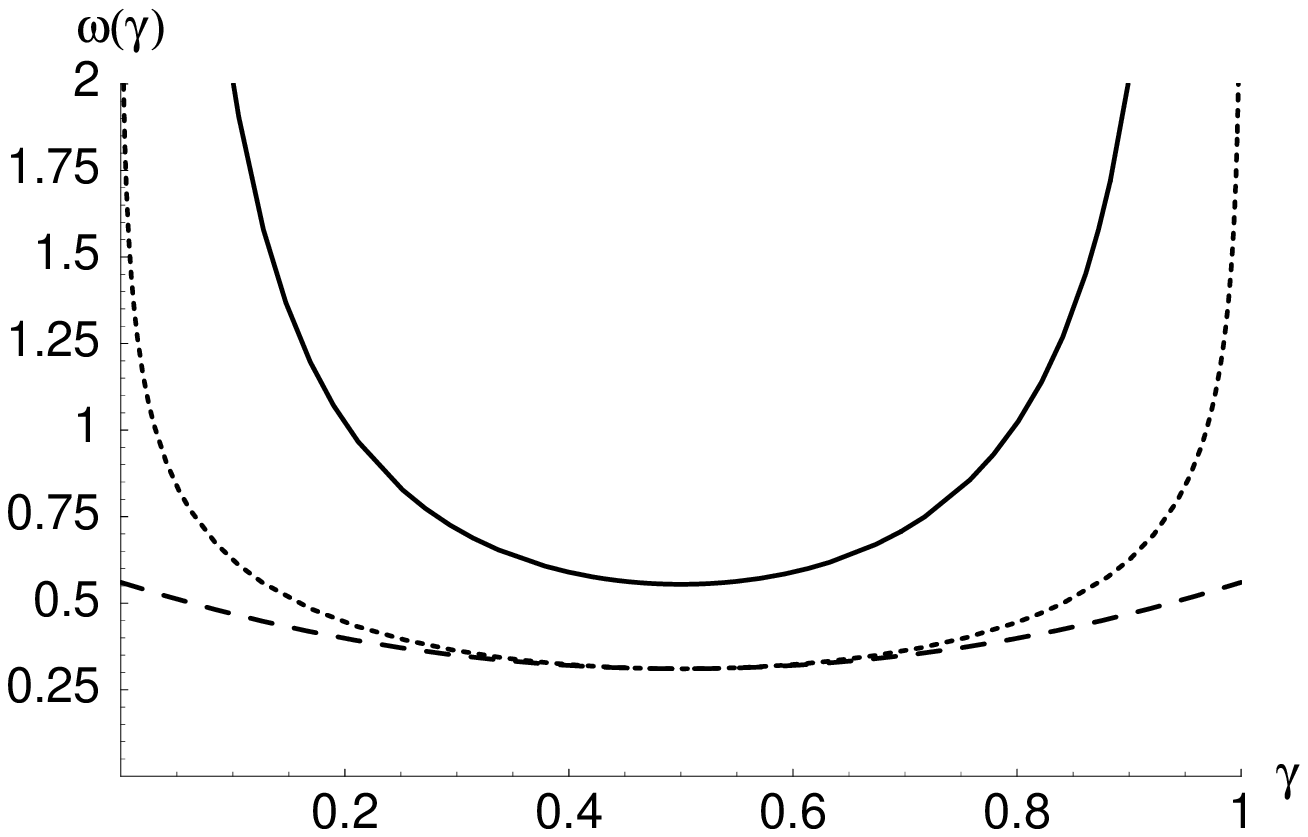,width=0.7\columnwidth}
\caption{Characteristic functions $\w(\g,\as=0.2)$. The solid line is LL BFKL, Eq.\ \protect\eq{chi0}, the dashed line is the kinematical constraint, Eq.\ \protect\eq{chi-CC}, and the dotted line is the rapidity veto, Eq.\ \protect\eq{chi-rapveto}, for $\eta=1.87$, which gives the same value for the pomeron intercept $\w(\frac{1}{2},\as)$ as the dashed line.\label{CC-fig}}
\end{figure}

In Fig.\ \ref{CC-fig} the solutions of the implicit equation for $\w(\g,\as)$ for a fixed $\as=0.2$ for the kernel \eq{chi-CC} is compared with the LL kernel $\x_0$. I also show the kernel \eq{chi-rapveto} with the rapidity veto for a choice of $\eta$ that gives the same pomeron intercept as the kinematical constraint.

The kernel \eq{chi-CC} above is symmetric with respect to the interchange $\g\to 1-\g$, just as the original LL BFKL kernel. It is thus written for a symmetric scale choice $s_0$ in the evolution variable $Y=\ln (s/s_0)$, were the scales on both sides of the ladder appear in $s_0$. In the present analysis we would like to make an asymmetric scale choice. In such a case the kinematical constraint should be formulated slightly differently \cite{CCS}, namely one writes the kernel as
\be
\x(\g,\w)=2\psi(1)-\psi\left(\g\right)-\psi\left(1-\g+\w\right),
\label{chi-CC-asy}
\ee
which is now asymmetric with respect to $\g\to 1-\g$. This form will be used below.

\subsection{Khoze--Martin--Ryskin--Stirling model}

Khoze, Martin, Ryskin and Stirling (KMRS) \cite{Khoze:2004hx} proposed a simpler model for the NLL corrected LL kernel, which is based on the approach of \cite{CCS} but with a fixed coupling. This kernel has also been used to estimate the effect of higher order corrections in diffractive $\gamma^*\gamma^*\to\rho\rho$ production \cite{EPSW} which provides a possibility to test it against the full NLL calculation \cite{Ivanov:2005gn}.

The characteristic function is expressed as
\be
\x(\g,\w) = \x_0(\g) + \as \x_1(\g,\w)
\label{durhamchi}
\ee
where $\x_0(\g)$ is the usual BFKL function \eq{chi0}. The correction piece $\x_1(\g,\w)$ is given by
\be
\as \x_1(\g,\w) =
\frac{1+ \w A_1(\w)}{\g} -
\frac{1}{\g} + 
\frac{1+ \w A_1(\w)}{1-\g+\w} -
\frac{1}{1-\g} -
\w \x_0^{\text{ht}}(\g), 
\ee
where $A_1(\w)$ is related to the Mellin transform of the DGLAP splitting function and $\x_0^{\text{ht}}$ is the higher twist part of $\x_0$,
\be
\x_0^{\text{ht}}(\g) = \x_0(\g) - \frac{1}{\g} - \frac{1}{1-\g}
= 2\psi(1) - \psi(1+\g) -\psi(2-\g).
\ee
We make the approximation that $n_F=0$ in the evolution, which gives $A_1(\w)\simeq -11/12$. It is possible to include the effects of quarks \cite{Khoze:2004hx} but they will be neglected here.

\begin{figure}[t]
\epsfig{file=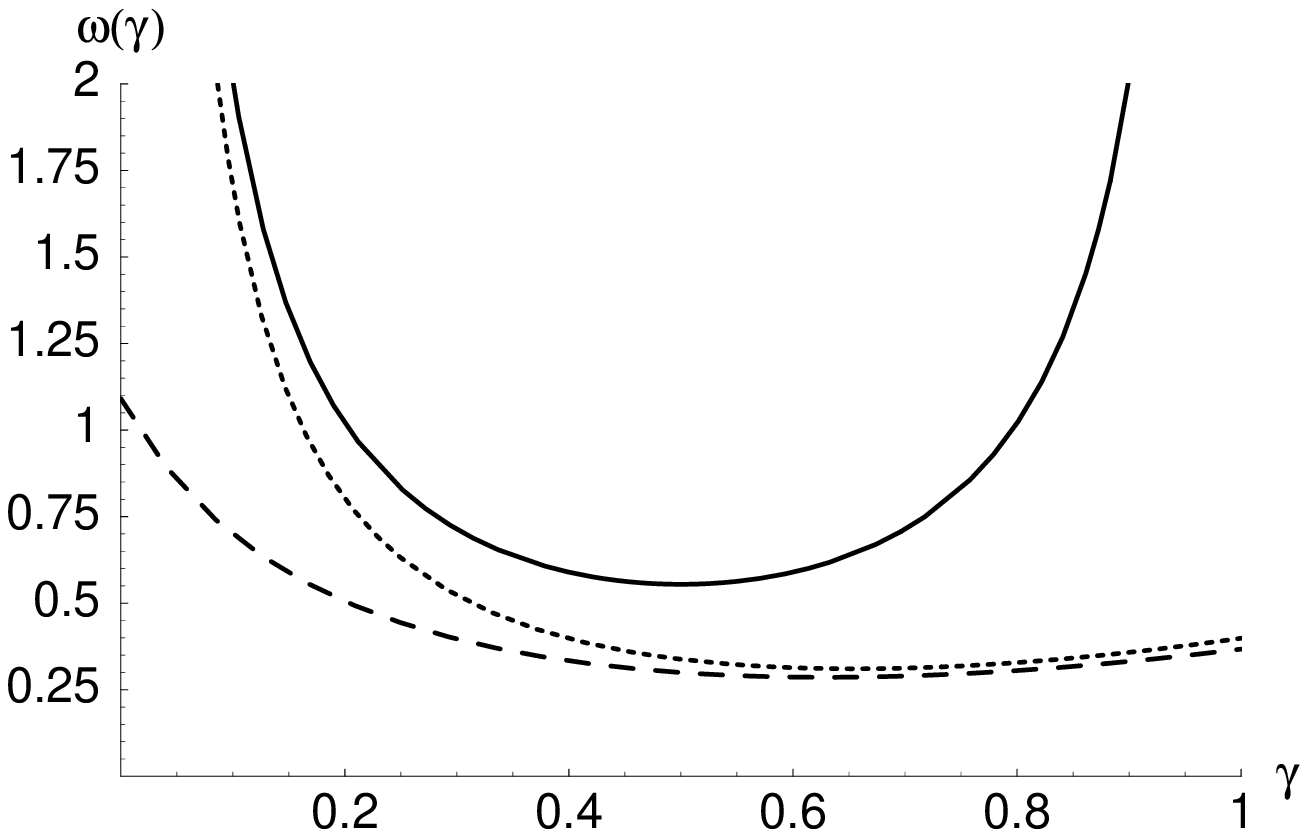,width=0.7\columnwidth}
\caption{Characteristic functions $\w(\g,\as=0.2)$. The solid line is LL BFKL, Eq.\ \protect\eq{chi0}, the dotted line is the asymmetrical kinematical constraint, Eq.\ \protect\eq{chi-CC-asy}, and the dashed line is the KMRS model \protect\eq{durhamchi}.\label{durham-fig1}}
\end{figure}

The characteristic kernel \eq{durhamchi} is shown in Fig.~\ref{durham-fig1} together with the LL BFKL kernel and the asymmetric consistency constraint. An important feature of the kernel  \eq{durhamchi} is that it has no pole at $\g=0$. Instead it approximately fulfills energy conservation through $\gamma(\w=1)\approx 0$ (exact energy conservation implies equality).

\subsection{Comparison of results}

Let us now compare the results obtained from the different suggested kernels. We will focus on two quantities, the derivative of the saturation scale and the reduced front shape.

\begin{figure}[th]
\epsfig{file=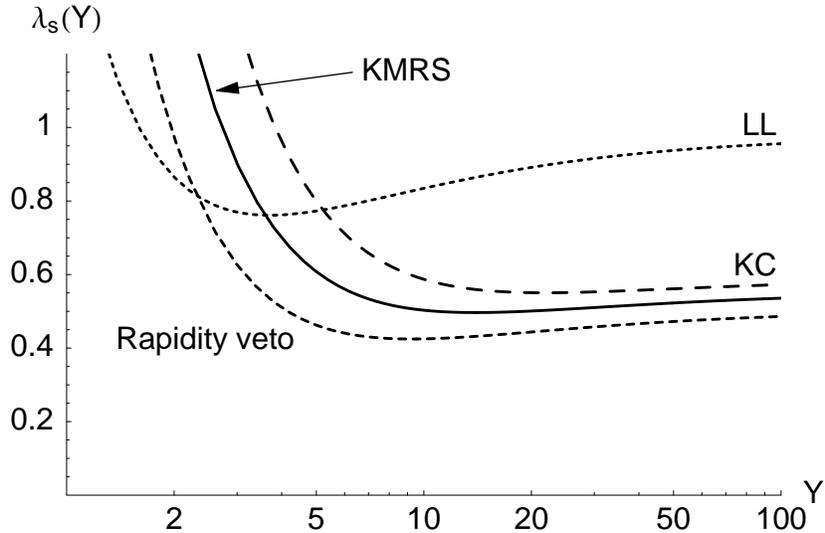,width=0.7\columnwidth}
\caption{Derivative of the saturation scale, $\lambda_s(Y)$, as defined in Eq.\ \protect\eq{satscder} for LL BFKL, the kinematical constraint (KC), the rapidity veto, and the KMRS kernel.\label{lambda-FC}}
\end{figure}

The logarithmic derivative 
\be
\lambda_s(Y) \equiv \frac{\partial \ln Q_s^2}{\partial Y}
\label{satscder}
\ee
gives a measure of how fast the fronts propagate toward large momenta and should asymptotically approach the constant value $v^*$. In Fig.\ \ref{lambda-FC}, the results for $\lambda_s(Y)$ obtained from the three different kernels are plotted together with the LL BFKL result. It is clear that the NLL corrections decrease $v^*$, but also that the approach to the asymptotics is somewhat smoothed out. This can be traced to the smaller values of the diffusion constants. For experimentally observable rapidities it seems that the velocity is a decreasing function of $Y$. Remember, however, that for smaller rapidities there can be appreciable effects coming from the non-universal terms that are not included here (see \cite{EGM} for numerical investigations of this).

\begin{figure}[ht]
\epsfig{file=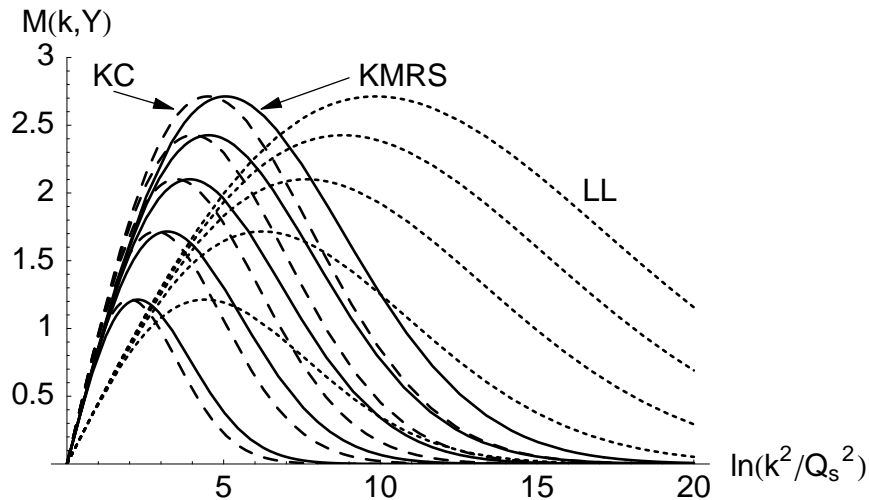,width=0.7\columnwidth}
\caption{Reduced front shape as defined in Eq.\ \protect\eq{redfront}, 
obtained from LL BFKL, the kinematical constraint, and the KMRS kernel, and 
plotted for rapidities $Y=2,4,6,8,10$. 
\label{redfront-FC}}
\end{figure}

The reduced front shape is obtained as the combination 
\begin{align}
{\cal M}(k,Y) &\equiv {\cal N}(k/Q_s(Y),Y) \times (k^2/Q_s^2(Y))^{\gamma_c}\\
&= \frac{1}{\sqrt{D}}
\ln\left(\frac{k^2}{Q_s^2(Y)}\right) 
\exp\left(-\frac{1}{4 D Y}\ln^2\left(\frac{k^2}{Q_s^2(Y)}\right) 
\right)
\label{redfront}
\end{align}
which removes the exponential part and quantifies the deviation from geometric scaling. For perfect geometric scaling the reduced front should not depend on the rapidity. In Fig.\ \ref{redfront-FC} the reduced fronts are plotted for the LL BFKL kernel, the consistency constraint and the KMRS model. The curves have widths which reflect the diffusion constants in the diffusive breaking of geometric scaling, and it is obvious that the RG-improved kernels lead to a slower diffusive breaking of geometric scaling.

\section{Diffusive differential equation approximation}\label{diffsection}\label{VI}

In \cite{MP1} the BK equation was approximated as a partial differential equation, which could be transformed into the FKPP equation by expanding the BFKL kernel to second order. The RG-improved BK equation may similarly be approximated as a differential equation, which will be a generalization of the FKPP equation. It is instructive to study this approximation since we then obtain closed analytical expressions and the structure of the modifications to the traveling wave equations may be easier appreciated.

The BFKL function $\chi$ can be Taylor expanded around the point $(\gamma_c,\omega_c)$ to second order,
\ba
\chi(\gamma,\omega) &\approx& 
\chi 
+ \xg (\gamma-\gc)
+ \xw (\omega-\omega_c) \nonumber\\
&+& \frac{1}{2} \left[
\xgg (\gamma-\gc)^2
+ \xww (\omega-\omega_c)^2 
+  2 \xgw 
(\gamma-\gc)(\omega-\omega_c)
\right]
\ea
where in this Section the subscript $c$ on $\x$ is omitted for simplicity of notation, so that
$\x = \chi(\gc,\omega_c)$,
$\xg = \partial_\gamma \chi(\gc,\omega_c)$,
$\xw = \partial_\omega \chi(\gc,\omega_c)$,
$\xgw = \partial_\gamma \partial_\omega\chi(\gc,\omega_c)$,
and so on. The BK equation for ${\cal N}={\cal N}(L,Y)$ in the diffusive approximation then takes the form
\ba
\dY {\cal N} &=& \as\left[ 
\frac{\xgg }{2} \, \dL^2
+ \frac{\xww}{2}\, \dY^2
+ (\xgg \gc - \xg + \xgw \wc)\, \dL 
- (\xww \wc - \xw + \xgw \gc)\, \dY
\right. \nonumber\\
&&\left.
- \xgw \,\dL\dY + \left( \x 
- \xg \gc - \xw \wc 
+ \frac{\xgg {{\gc }^2}}{2} 
+ \xgw \gc \wc 
+ \frac{\xww {{\wc }^2}}{2}
\right)
\right] {\cal N} - \as {\cal N}^2
\label{BK-NLL-diffusive}
\ea
which at first glance looks more symmetric in $L$ and $Y$ than the LL approximation.
The equation contains a cross term derivative and a term proportional to $\dL$, but the variable change
\be
\begin{cases}
t = A Y + B L \\
x = C Y + D L
\label{firstchange}
\end{cases}
\ee
gives an equation on the form
\be
\tau\, \dt^2 u + \dt u
= \dx^2 u +  u -  u^2. \label{ext-KPP}
\ee
The explicit expressions for the constants $\tau$, $A$, $B$, $C$, and $D$ in terms of $\x,\xg,\xw$ etc.\ are not very illuminating and are given in the Appendix. The function $u$ is defined as
\ba
u(x,t) &=& \text{const} \times  
{\cal N}\left(L=\frac{A x -C t}{AD-BC},Y=\frac{D t- B x}{AD-BC}\right).
\ea
Remember that all the coefficients are just pure numbers once the saddle point values $\gc$ and $\wc$ have been found.

Eq.\ \eq{ext-KPP} is an extended FKPP equation with a higher time derivative. Note that in LL BFKL the evolution occurs in rapidity, since the time variable $t \propto Y$, but in the present case the evolution instead occurs in $t = A Y + B L$. The $\omega$-dependent kernel thus mixes evolution in rapidity and momentum.

We can now analyze Eq.\ \eq{ext-KPP} in the same way as we analyzed the full linearized BK equation in the LL or NLL cases. The result is
\ba
v^* &=& \frac{2}{\sqrt{1+ 4\tau}}\\
\lambda^* &=& \sqrt{1+ 4\tau}\\
D &=& \frac{1}{(1+ 4\tau)^2}.
\ea
This is to be compared with the results for the first-order in time FKPP equation, 
$v^*=2, \lambda^*=1, D=1$. Taking into account the transformation back to the variables $L$ and $Y$ we again see that the asymptotic velocity is reduced compared to the LL value. 

It turns out, however,  to be much simpler to accept a ``convection term'' proportional to $\dx$ in the equation. This makes it possible to keep the $t \propto Y$ property and the change of variables is much simpler since the only constraint is to get rid of the mixed derivative term instead of getting rid of several terms. In fact, such a convection term will only change the velocity by an additive constant.

Thus, if we instead in Eq.\ \eq{BK-NLL-diffusive} make the variable change
\be
\begin{cases}
t = A Y  \\
x = C Y + D L,
\label{secondchange}
\end{cases}
\ee
we get the transformed equation
\be
\tau\, \dt^2 u + \dt u
= \dx^2 u + \sigma \, \dx u + u -  u^2. \label{ext2-KPP}
\ee
with
\ba
u(x,t) &=&  \text{const} \times  
{\cal N}\left(L=\frac{x}{D} -\frac{C t}{A D},Y=\frac{t}{A}\right).
\ea
The coefficients $\tau$, $\sigma$, $A$, $C$, and $D$ are again given in the Appendix.
This equation has an additional linear derivative in $x$, with a ``convection coefficient'' $\sigma$. With this change of variables we therefore see the problem in a different way; the evolution in time is now evolution in rapidity  as in LL instead of in rapidity and momentum as with the earlier variable change. There is a higher order time derivative slowing down the spreading, as well as a convection term damping the wave.
Unfortunately the expressions for $v^*$, $\lambda^*$ and $D$ are now quite complicated and I choose to not write them here.

The interesting points here are that when passing to RG-improved kernels, the spreading velocity decreases, and the equation becomes more symmetric in $L$ and $Y$. In terms of linearized partial differential equation approximations, the LL BFKL equation is first order in time and second order in space while the DGLAP equation is first order in space and second order in time. The RG-improved equation which contains higher order corrections is instead second order in both time and space. This duality is explicit in Eqs.\ \eq{ext-KPP} and \eq{ext2-KPP}, reminding one of the approach of Ref.~\cite{ABF}.

\section{Summary and outlook}

I have proposed a simple way to use the traveling wave framework to compute the energy dependence of the saturation scale for RG-improved BFKL kernels with fixed QCD coupling. This completes the similar computation done for running coupling in \cite{PS}. Higher order corrections are found to suppress both the increase with energy of the saturation scale and the speed of the diffusive spread of the front that breaks geometric scaling. The results obtained from three of the different kernels proposed in the literature are somewhat different in size. I have also studied the differential equation approximation of the evolution equation and found that the RG-resummation qualitatively changes the obtained equations.

The study of RG-improvements to the BK equation at fixed or running coupling is the first step in understanding higher order corrections to saturation equations. Recent studies of how to rigorously include a running coupling into the BK and Balitsky--JIMWLK equations are not conclusive \cite{running}, and the full NLL corrections to the BK equations have not been computed. It is not clear if the linearized version of the NLL BK equation would be the same as the NLL BFKL equation, but it is a reasonable assumption. Therefore, at this time, it is useful to consider the effect of RG-resummed kernels on the solutions of the BK equation. 

It would be interesting to compare the results obtained in this paper to numerical simulations for various kernels. The results could also be used in comparing to experimental data along the lines of e.g.~\cite{Iancu:2003ge}. Another interesting question is to ask what happens when fluctuations~\cite{Iancu:2004es} are included in the picture.

\begin{acknowledgments}
I am very grateful to St\'ephane Munier for support, discussions and helpful suggestions. I also want to thank him and Robi Peschanski for useful comments on the manuscript.
\end{acknowledgments}

\appendix*
\section{Variable changes for diffusive approximation}

The first change of variables, Eq.\ \eq{firstchange}, is defined by
\be
\begin{cases}
t = A Y + B L \\
x = C Y + D L
\end{cases}
\ee
where
\ba
A &=&  -\dfrac{\as \xgw { }^2 \wc + \xw \as  \xgg-\xgg-\as\xg \xgw 
-\as \xgg \xww \wc }
{\as   {\xgw}{ }^2 \gc  - \as \xw \xgw + \xgw+
\as\xg \xww  -\as\xgg \xww   \gc } B \\
D &=&  \dfrac{\xw - \xgw  \gc - \xww  \wc - 1/\as}{ \xg-\xgg \gc - \xgw  \wc} C,
\ea
and where $B$ and $C$ are given by $B =  \beta_4/\beta_2$ and $C^2 =  \beta_4/\beta_3$ with
\ba
\beta_1 &=&  
-\frac{1}{2}\,  \left(\frac{\xww\, {{(\xgw\, \as \, (\xg-\xgw\,
\wc )+\xgg\, (-\xw\, \as +\xww\, \wc \, \as +1))}^2}}{{(\as \, \gc \,
{{\xgw}{}^2}-\xw\, \as \, \xgw+\xgw+\xww\, \as \, (\xg-\xgg\,
\gc ))}^2}+ \right. \nonumber\\ 
&& \left.
\xgg+\frac{2\, \xgw\, (\xgw\, \as \, (\xgw\, \wc -\xg)+\xgg\,
(\xw\, \as -\xww\, \wc \, \as -1))}{\as \, \gc \, {{\xgw}{}^2}-\xw\,
\as \, \xgw+\xgw+\xww\, \as \, (\xg-\xgg\, \gc )}\right)
\\
\beta_2 &=&  
 \left({{\xgg}{}^2}\, \xww\, {{\as }^2}\, {{\gc }^2}+\xgg\, ({{\xw}{}^2}\,
{{\as }^2}-{{\xgw}{}^2}\, {{\gc }^2}\, {{\as }^2}+{{\xww}^2}\, {{\wc }{}^2}\, {{\as }^2}- \right. \nonumber\\
&& 2\, \xg\, \xww\, \gc \, {{\as }^2}+2\, \xgw\, \xww\,
\gc \, \wc \, {{\as }^2}+2\, \xww\, \wc \, \as -2\, \xw\, (\xww\,
\as \, \wc +1)\, \as +1)+ \nonumber \\
&& \left. \as \, (\xg-\xgw\, \wc )\, (2\, \as \, \gc \, {{\xgw}{}^2}+(-2\,
\xw\, \as +\xww\, \wc \, \as +2)\, \xgw+\xg\, \xww\,
\as ) \right)   \nonumber\\
&&  / \left( \as \, (\as \, \gc \, {{\xgw}{}^2}-\xw\, \as \, \xgw+\xgw+\xww\,
\as \, (\xg-\xgg\, \gc )) \right)
\\
\beta_3 &=&   
\frac{1}{2}\, \left(\frac{\xgg\, {{(-\xw\, \as +\xgw\, \gc \, \as
+\xww\, \wc \, \as +1)}^2}}{{{\as }^2}\, {{(-\xg+\xgg\, \gc +\xgw\,
\wc )}^2}}+\nonumber\right.\\
&&\left.\frac{2\, \xgw\, (-\xw\, \as +\xgw\, \gc \, \as +\xww\,
\wc \, \as +1)}{\as \, (\xg-\xgg\, \gc -\xgw\, \wc )}+\xww\right)
\\
\beta_4 &=&  
\frac{1}{2} \left(\xgg\, \gc^2-2\, \xg\, \gc +2\, \xgw\, \wc \, \gc
+\xww\, {\wc {}^2}+2\, \x-2\, \xw\, \wc \right).
\ea
Finally, $\tau$ is given by $\tau = \beta_1 \beta_4/\beta_2^2$.
 
The second variable change, Eq.\ \eq{secondchange}, is defined by the somewhat simpler form
\be
\begin{cases}
t = A Y  \\
x = C Y + D L,
\end{cases}
\ee
where 
\be
A = \frac{\beta_5}{\beta_2}, \quad
C = D \frac{\xgw}{\xww}, \quad 
D^2 = \frac{\beta_5}{\beta_3}, 
\ee
and where the coefficients $\beta_i$ are
\ba
\beta_1 &=& -\frac{\xww}{2}  \nonumber \\
\beta_2 &=& \xww \, \wc  + \xgw \, \gc -\xw  + 1/\as \nonumber \\
\beta_3 &=&  \frac{\xgg\, \xww-{\xgw{}^2}}{2\, \xww} \\
\beta_4 &=&  \frac{1}{\xww}\, \left(\xgg \, \xww\, \gc - \xgw{}^2 \, \gc +\xw\,  \xgw - \xg\, \xww - \frac{\xgw}{\as}    \right) \nonumber \\
\beta_5 &=& 
\frac{\xgg\, {\gc^2}}{2} + \xgw\, \wc \, \gc  + \frac{\xww\,{\wc^2}}{2} - \xg\, \gc - \xw\, \wc  + \chi.
\nonumber 
\ea
We also have $\tau = {\beta_1 \beta_5}/{\beta_2^2}$ and
$\sigma = \sqrt{{\beta_4^2}/{\beta_3 \beta_5}}$.


\end{document}